\documentclass[journal,twoside,web]{ieeecolor}
\usepackage{generic}
\usepackage{cite}
\usepackage{amsmath,amssymb,amsfonts}
\usepackage{algorithmic}
\usepackage{graphicx}
\usepackage{algorithm,algorithmic}
\usepackage{booktabs} 
\usepackage{makecell} 
\usepackage{textcomp}
\def\BibTeX{{\rm B\kern-.05em{\sc i\kern-.025em b}\kern-.08em
		T\kern-.1667em\lower.7ex\hbox{E}\kern-.125emX}}
\makeatletter
\def\thebibliography#1{\section*{References}%
	\footnotesize\rmfamily 
	\list{\@biblabel{\arabic{enumi}}}{\def\makelabel##1{##1}%
		\settowidth\labelwidth{\@biblabel{#1}}%
		\leftmargin\labelwidth \advance\leftmargin\labelsep
		\usecounter{enumi}}%
	\def\newblock{\hskip .11em plus .33em minus .07em}%
	\sloppy\clubpenalty4000\widowpenalty4000
	\sfcode`.=1000\relax}
\makeatother

\begin{document}
	\title{Dual-Correction Physics-Informed Neural Networks for Hemodynamic Reconstruction from Sparse Data}
	\author{Jingtai Song, Qinsheng Zhu, Xiaodong Xing,  Yufeng Tang, Zhiyun Zhang, Xianwen Zhang, and Hao Wu
		\thanks{Manuscript received XXXX; revised XXXX; accepted XXXX. Date of publication XXXX; date of current version XXXX. This work was supported in part by the Natural Science Foundation of Xinjiang Uygur Autonomous Region under Grant 2024D01A17; in part by the Chengdu Key Research and Development Program under Grant 2025-YF08-00109-GX; in part by the Research Fund of Sichuan Medical Association under Grant 2025RY15; and in part by the Clinical Key Special Fund of Mianyang Central Hospital under Grant 2025LC004. (Corresponding authors: Qinsheng Zhu.)}
		\thanks{This study retrospectively utilized anonymized clinical imaging data. The authors confirm that the data usage protocols were approved by the Institutional Review Board (IRB) of Mianyang Central Hospital, and the requirement for informed consent was waived.}
		\thanks{Jingtai Song, Qinsheng Zhu, and Hao Wu are with the School of Physics, University of Electronic Science and Technology of China, Chengdu 611731, China. Qinsheng Zhu is also with the Kashgar Institute of Electronic Information Industry Technology, Kashgar 844000, China (e-mail: zhuqinsheng@uestc.edu.cn).}
		\thanks{Xiaodong Xing is with School of Quantum Information Future Technology, Henan University, Zhengzhou 450046, China; Henan Key Laboratory of Quantum Materials and Quantum Energy, Henan University, Zhengzhou 450046, China; Institute of Quantum Materials and Physics, Henan Academy of Sciences, Zhengzhou, 450046 China .}
		\thanks{Yufeng Tang, Zhiyun Zhang, and Xianwen Zhang are with the Department of Neurology, Mianyang Central Hospital, School of Medicine, University of Electronic Science and Technology of China, Mianyang 621000, China. Yufeng Tang is also with the Sichuan Provincial Engineering Research Center of Nuclear Medical Equipment Translation and Application, Mianyang Central Hospital, Mianyang 621000, China. Xianwen Zhang is also with the Sichuan Clinical Research Center for Radiation and Therapy, Mianyang Central Hospital, Mianyang 621000, China .}}
	
	\maketitle
	
	\begin{abstract}
		Quantifying hemodynamics in the curved segments of the intracranial internal carotid artery is a core challenge in diagnosing vascular stenosis. Conventional full-field imaging, such as 4D Flow MRI, is costly and difficult to widely promote. Meanwhile, reconstructing full-field fluid information from easily accessible and non-invasive sparse measurement data (such as transcranial Doppler ultrasound/computed tomography angiography) is essentially a highly challenging ill-posed inverse problem. To overcome the severe optimization difficulties and generalization failures of conventional physics-informed neural networks (PINNs) in highly tortuous geometries, we propose a dual-correction physics-informed neural network (DCP-INN) framework taking into account a causal decoupling strategy. The proposed DCP-INN model utilizes a diamond-shaped main network to capture low-frequency trends in physical evolution, and employs a parallel wide-deep correction network to compensate for high-frequency residuals resulting from complex geometric shapes. Furthermore, the framework introduces a high-order physical loss function based on Taylor expansion to enhance local continuity under extremely sparse data constraints. To validate the proposed method, we performed computational evaluations on realistic vascular geometries with significant tortuosity. The results demonstrate that the method effectively mitigates optimization challenges and significantly reduces flow field reconstruction error. This study not only achieves physically credible and robust flow field reconstruction in complex morphologies but also provides a highly promising algorithmic foundation for building low-cost, high-resolution personalized cardiovascular digital twins in future.
	\end{abstract}
	
	
	\section{Introduction}
	\label{sec:introduction}
	
	\IEEEPARstart{I}{schemic} stroke caused by internal carotid artery stenosis (ICAS) is one of the leading causes of death and long-term disability worldwide, and its precise prevention and treatment constitute a core challenge in modern neuromedicine. Currently, clinical risk assessment for ICAS has shifted from relying solely on static anatomical stenosis rates to the comprehensive quantification of hemodynamic parameters such as trans-stenotic pressure gradients and complete spatiotemporal velocity fields. However, conventional full-field imaging (e.g., 4D Flow MRI) is costly, and high-resolution CTA lacks flow velocity features. Consequently, clinical practice highly relies on portable transcranial Doppler (TCD) ultrasound for assessment. Nonetheless, TCD is essentially a single-point measurement limited by individual ``acoustic windows.'' Despite its clinical utility, TCD remains fundamentally limited in its ability to resolve continuous spatial flow fields within highly tortuous vascular structures, such as the intracranial carotid siphon. This spatial incompleteness introduces considerable uncertainty into hemodynamic assessment, potentially compromising the reliability of clinical decision-making. Consequently, accurate reconstruction of full-field hemodynamic evolution in ICAS patients from sparse, non-invasive TCD measurements represents a pressing and largely unresolved clinical challenge.~\cite{10.1016/j.media.2009.07.011, 10.1016/j.cmpb.2018.02.001, 10.1063/5.0259296}.
	
	To address the highly ill-posed problem of reconstructing continuous full-field hemodynamics from extremely sparse clinical observations, the core challenge lies in achieving precise fusion between physical governing equations and the limited measurement data. Traditional computational fluid dynamics (CFD) relying on data assimilation or Kalman filtering suffers from tedious mesh generation and extreme sensitivity to boundary conditions~\cite{10.1002/cnm.3639}. Conversely, pure data-driven machine learning (e.g., standard deep neural networks) lacks physical constraints, risking severe overfitting on sparse data. These limitations compel a pivot toward mesh-free physics-informed deep learning paradigms capable of integrating physical priors with data flexibility.
	
	To achieve the ultimate goal of robust, mesh-free hemodynamic reconstruction from sparse clinical data, physics-informed neural networks (PINNs) provide a highly promising solution~\cite{10.1016/j.jcp.2018.10.045, 10.1109/tmi.2022.3161653, 10.1109/tmi.2025.3587636}. By directly embedding physical governing equations (e.g., 1D hemodynamic models) into the loss function of the neural network as regularization terms, PINNs can learn continuous spatiotemporal representations of physical fields from sparse, noisy data within a mesh-free framework. This circumvents the multiple iterative solutions and prior assumptions on error statistics required by traditional data assimilation methods (like Kalman filtering). Although PINNs perform excellently in regular vessels, their reconstruction error surges to over 40\% when generalized to real clinical vessels with highly tortuous morphologies. This study assumes that the root cause of this failure lies in the fundamental contradiction between the high-frequency nonlinearities induced by complex geometries and the inherent ``spectral bias'' of deep networks. Drastic changes in the cross-sectional area induce rich high-frequency components, whereas standard multi-layer perceptrons naturally prioritize learning low-frequency trends. This forces standard PINNs into a dilemma: either ignoring flow perturbations to satisfy global smoothness constraints (physical underfitting) or generating non-physical high-frequency oscillations to forcefully fit sparse probes (local overfitting).

    Synthesizing the aforementioned clinical realities and algorithmic challenges, the core task of this study can be formulated as a severely data-deficient ill-posed inverse problem. Mathematically, the idea is to identify a highly nonlinear mapping operator $\mathcal{G}$ that takes two inputs: the geometric descriptor $A_0(x)$ , extracted and dimensionality-reduced from patient-specific 3D imaging data, and the sparse velocity time-series $u(t)$, acquired at a limited number of discrete measurement probes. The operator $\mathcal{G}$ is then required to produce a continuous, high-resolution full-field blood velocity solution $u(x,t)$ over the entire spatiotemporal computational domain.
	
	To resolve this optimization conflict co-induced by ``spectral bias'' and ``geometric nonlinearity'' and successfully approximate the aforementioned complex mapping operator $\mathcal{G}$, this paper proposes a dual-correction physics-informed neural network (DCP-INN) for sparse data hemodynamic reconstruction. The main contributions are as follows: 1) A decoupled ``main + correction network'' architecture is proposed, where the main network captures low-frequency flow field trends, and the parallel correction network compensates for geometry-induced high-frequency residuals. 2) A Taylor expansion-based physical loss (Taylor Loss) is innovatively introduced into 1D hemodynamic reconstruction, serving as a strong neighborhood constraint to significantly enhance physical fidelity near sparse data points. 3) The superiority of an asymmetric ``diamond-shaped'' network topology in balancing feature extraction and information regression is revealed. 4) Computational validations on realistic vascular geometries derived from clinical imaging demonstrate that the proposed method significantly reduces reconstruction errors and exhibits strong robustness in resolving complex geometric bottlenecks.

    The structure of this article is organized as follows. Section II details the mathematical formulation and the proposed DCP-INN methodology. Section III presents the data preparation, evaluation metrics, and the flow field reconstruction results. Section IV discusses the clinical implications and limitations of our method. Finally, Section V concludes this study.

	\section{Methods}
	\label{sec:methods}
	
\begin{figure*}[!t]
		\centering
		\includegraphics[width=0.85\textwidth]{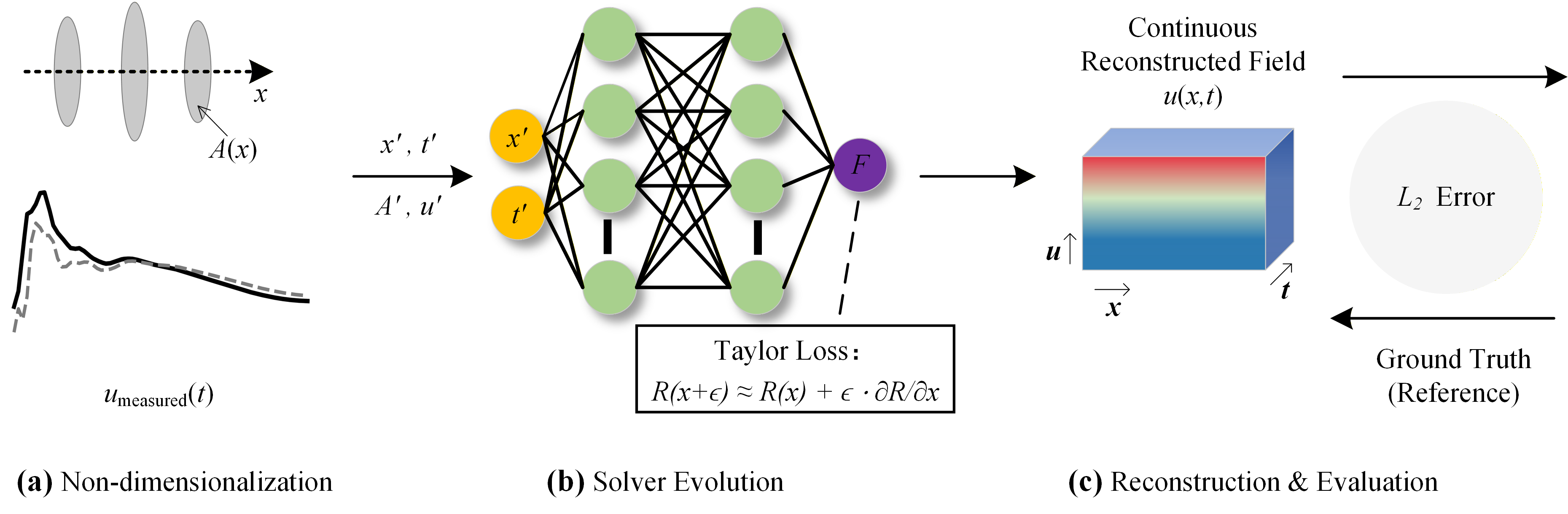}
		\caption{Overview of the computational workflow and methodological evolution. (a) Non-dimensionalization: The geometric function $A(x)$ and sparse probe measurements are non-dimensionalized prior to integration into the computational framework. (b) Solver Evolution: The Taylor loss is introduced upon the fundamental physical constraints, with the learning and solving processes executed via the PINN. (c) Reconstruction and Evaluation: The solver outputs a high-resolution, continuous reconstructed flow field, which is compared against the ground truth via the $L_2$ error metric.}
		\label{fig:method_evolution}
	\end{figure*}

	The methodology presents the fundamental challenge of accurately reconstructing complex hemodynamics from sparse data. We start with establishing the mathematical formulation of this inverse problem (Section A). Building on this foundation, we further introduce and evaluate two components designed to enhance the performance of generic PINNs (Section B). Subsequently, we elucidate the optimization convergence limitations encountered by these enhancement methods when applied to complex geometries with high curvature. This analysis provides the theoretical rationale for the core contribution of this work: a novel Double Correction Physics-Informed Neural Network (DCP-INN) architecture (Section C). Finally, we summarize the network design principles distilled through systematic tuning and provides the comprehensive implementation details and evaluation metrics of the final model (Section D). As illustrated in Fig. \ref{fig:method_evolution}, the methodological workflow is partitioned into three core stages: 1) non-dimensionalization of geometric priors and sparse observations; 2) solver evolution integrated with physical constraints; and 3) reconstruction and evaluation of the high-resolution flow field. Within this module, we not only introduce the illustrated Taylor loss upon the fundamental physical constraints to enforce strong local gradient constraints, but also build upon this foundation to ultimately derive the dual-correction architecture designed to decouple the low-frequency baseline from high-frequency residuals. The entire workflow ultimately culminates in the continuous reconstruction of the full-spatiotemporal-resolution flow field, which is quantitatively benchmarked against the ground truth via the global $L_2$ error metric.
	
	\subsection{Mathematical Formulation of Physics-Informed Neural Networks}
	
	\subsubsection{Non-dimensionalized Governing Equations}
	The fluid computational domains in this study are directly extracted from authentic clinical imaging data. By employing the SimVascular software for 3D anatomical reconstruction and 1D geometric centerline abstraction, we accurately calculate the cross-sectional area function $A(x)$ along the path. The computational domains involved in our study originate from two authentic vessels with distinct geometric complexities, as illustrated in Fig. \ref{fig:domain}.  Vessel 0 represents a relatively regular vessel segment with low curvature; whereas Vessel 4, serving as the core challenge of this study, is a highly tortuous and complex vessel whose curvature variations introduce a nonlinear testing environment for the model. Within these specific geometric domains, the fluid dynamics are governed by the classical one-dimensional incompressible Navier-Stokes mass and momentum conservation equations~\cite{10.1016/j.cmpb.2024.108427, arxiv-1911.08655}. To formulate a well-posed mathematical model, we introduce an algebraic constitutive state equation based on an exponential function to accurately correlate the internal pressure $p$ with the instantaneous area $A$:
	
	\begin{figure}[!t]
		\centering
		\includegraphics[width=0.85\columnwidth]{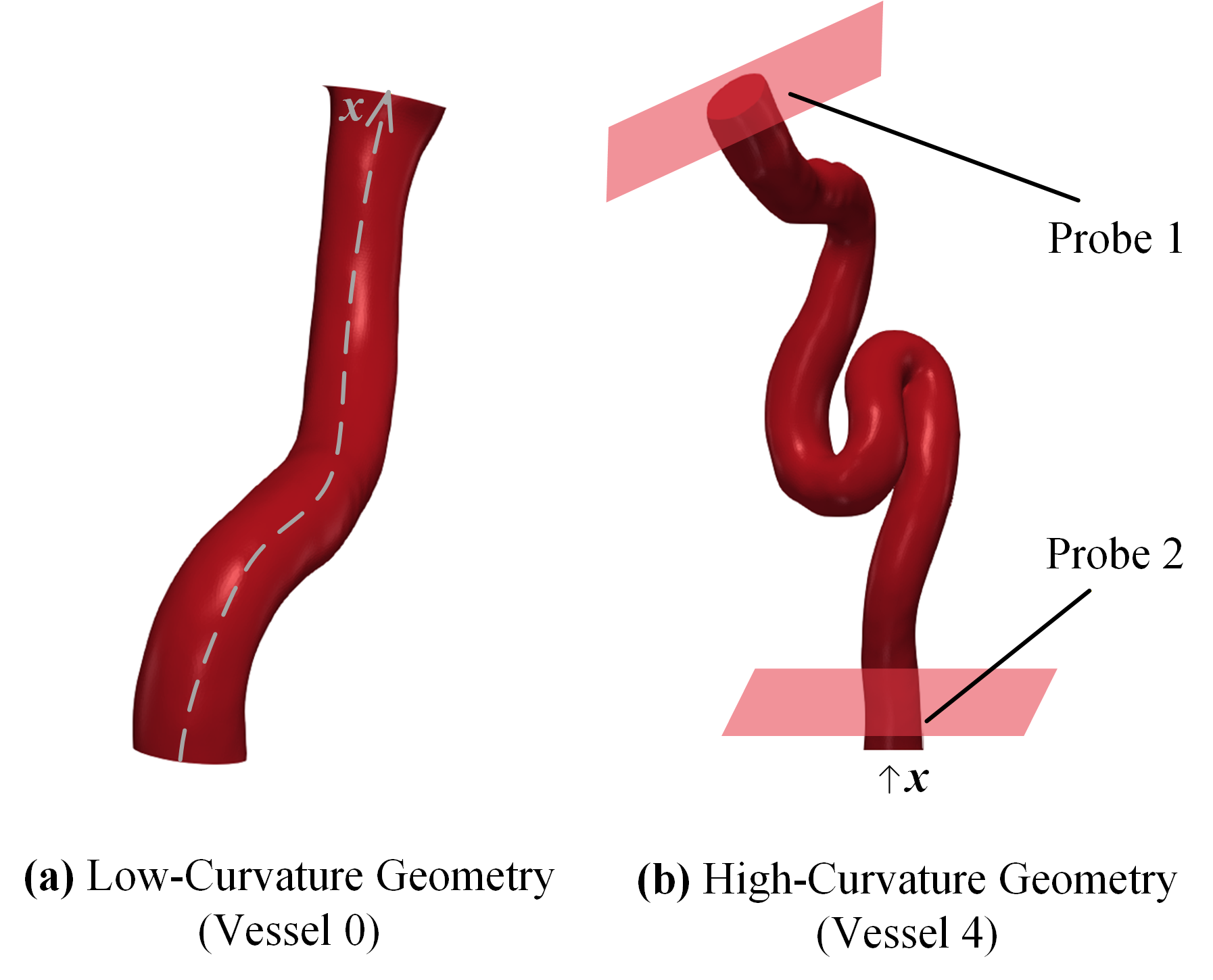}
		\caption{Computational domain and sparse data problem setup. This figure illustrates two realistic vessel models with distinct geometric complexities used in this study. (a) A relatively straight vascular segment (Vessel 0), where the white dashed line indicates the 1D spatial axis $x$. (b) A highly tortuous vascular segment (Vessel 4), where the two planes denote the spatial probe locations (Probe 1 and Probe 2) utilized for extracting sparse boundary data.}
		\label{fig:domain}
	\end{figure}
	
	\begin{equation}
		p(A) = p_0 + K_1 \exp\left(K_2 \frac{A}{A_0}\right)
		\label{eq:state}
	\end{equation}
    
    Here, $A_0$ and $p_0$ denote the reference cross-sectional area and reference pressure, respectively, while $K_1$ and $K_2$ represent empirical material constants that characterize the elasticity and stiffness of the vessel wall.
    
	To construct a numerically stable optimization problem that can be effectively learned by neural networks, we must non-dimensionalize the aforementioned physical prior equations. We define the dimensionless time ($t'$), spatial coordinate ($x'$), cross-sectional area ($A'$), and blood flow velocity ($V'$) as follows:
	
	\begin{equation}
		t' = \frac{t}{T_{\text{ref}}}, \quad x' = \frac{x}{L_{\text{ref}}}, \quad A' = \frac{A}{A_{\text{ref}}}, \quad V' = \frac{V}{V_{\text{ref}}}
		\label{eq:nondim_vars}
	\end{equation}
	
	where $T_{\text{ref}}$, $L_{\text{ref}}$, $A_{\text{ref}}$, and $V_{\text{ref}}$ denote the characteristic reference scales for their respective physical quantities. By substituting these dimensionless variables into the original equations, the governing equations can be rewritten in the following form:
	
	\begin{equation}
		\frac{\partial A'}{\partial t'} + c_1 \frac{\partial (A' V')}{\partial x'} = 0
		\label{eq:nondim_mass}
	\end{equation}
	
	\begin{equation}
		\frac{\partial V'}{\partial t'} + c_1 V' \frac{\partial V'}{\partial x'} + c_2 \frac{\partial p}{\partial x'} = 0
		\label{eq:nondim_momentum}
	\end{equation}
	
	Here, the factor $c_1=(T_{\text{ref}}\cdot V_{\text{ref}})/L_{\text{ref}}$ and $c_2=T_{\text{ref}}/(\rho \cdot V_{\text{ref}})$. The partial derivative of the pressure term is analytically computed via the chain rule combined with the equation of state, which is seamlessly integrated into the automatic differentiation computation graph.
	
	\subsubsection{Baseline PINN Formulation}
	The dimensionless flow variables $[A'(x',t'), V'(x',t')]$ are parameterized by a fully connected neural network $\text{NN}(x',t';\theta)$. We optimize the network parameters $\theta$ by minimizing a baseline total loss function, constructed as a weighted composition of a data fidelity term $\mathcal{L}_{\text{data}}$ and a physics constraint term $\mathcal{L}_{\text{phys}}$:
	
	\begin{equation}
		\mathcal{L}_{\text{baseline}} = W_{\text{data}} \mathcal{L}_{\text{data}} + W_{\text{phys}} \mathcal{L}_{\text{phys}}
		\label{eq:loss_baseline}
	\end{equation}
	
	The first component is the data fidelity loss $L_{\text{data}}$, which quantifies the discrepancy between the network predictions and the sparse, ground-truth observational data. It is composed of three Mean Squared Error (MSE) terms, corresponding to the initial condition and the boundary conditions at two probe monitoring points, respectively:
	
	\begin{equation}
		\begin{aligned}
			\mathcal{L}_{\text{data}} &= \frac{1}{N_{\text{ic}}} \sum_{i=1}^{N_{\text{ic}}} |A'(x_i,0) - A'_{\text{true},i}|^2 \\
			&\quad + \frac{1}{N_{\text{bc},A}} \sum_{j=1}^{N_{\text{bc},A}} |A'(x_{\text{probe}},t_j) - A'_{\text{true},j}|^2 \\
			&\quad + \frac{1}{N_{\text{bc},V}} \sum_{k=1}^{N_{\text{bc},V}} |V'(x_{\text{probe}},t_k) - V'_{\text{true},k}|^2
		\end{aligned}
		\label{eq:loss_data}
	\end{equation}
	
	where $N_{\text{ic}}$, $N_{\text{bc},A}$, and $N_{\text{bc},V}$ denote the number of sample points for the initial condition, area boundary condition, and velocity boundary condition, respectively.
	
	The second component, the physics constraint loss $L_{\text{phys}}$, acts to enforce the network predictions to satisfy the dimensionless governing equations across the entire spatiotemporal domain. To this end, we first define the physical residuals corresponding to the continuity equation ($R_{\text{cont}}$) and the momentum equation ($R_{\text{mom}}$):
	
	\begin{equation}
		R_{\text{cont}}(A', V') = \frac{\partial A'}{\partial t'} + c_1 \frac{\partial (A' V')}{\partial x'}
		\label{eq:res_cont}
	\end{equation}
	
	\begin{equation}
		R_{\text{mom}}(A', V') = \frac{\partial V'}{\partial t'} + c_1 V' \frac{\partial V'}{\partial x'} + c_2 \frac{\partial p}{\partial x'}
		\label{eq:res_mom}
	\end{equation}
	
	Here, all partial derivative terms (e.g., $\partial A'/\partial t'$, $\partial A'/\partial x'$) are precisely computed via the back-propagation and Automatic Differentiation (AD) mechanisms of the neural network~\cite{arxiv-1502.05767, 10.1016/j.cpc.2025.109569}. $L_{\text{phys}}$ is finally defined as the sum of the MSE of these two residuals evaluated at $N_c$ collocation points randomly sampled from the spatiotemporal domain $[0,1]\times[0,1]$:
	
	\begin{equation}
		\mathcal{L}_{\text{phys}} = \frac{1}{N_c} \sum_{i=1}^{N_c} \left( R_{\text{cont}}(x_i, t_i)^2 + R_{\text{mom}}(x_i, t_i)^2 \right)
		\label{eq:loss_phys}
	\end{equation}
	
	This $L_{\text{baseline}}$ loss function represents a standard PINN framework for solving this problem and serves as the performance benchmark for all our subsequent methodological enhancements.
	
	\subsection{Loss Function Design and Optimization Strategy}
	This section aims to explore algorithmic-level enhancement strategies to improve the physical consistency and convergence stability of the solver under sparse data constraints. Specifically, we introduce a higher-order derivative-based geometric constraint mechanism (Section B.1) and a dynamic loss weighting strategy (Section B.2), and systematically evaluate the effectiveness of these generic components in solving complex hemodynamic inverse problems.
	
	\subsubsection{Gradient-Enhanced Physics Loss}
	Under extremely sparse constraints, the standard physics loss $\mathcal{L}_{\text{phys}}$ acts as a weak ``pointwise constraint'' ($R(x_i) \approx 0$), rendering the model prone to parasitic high-frequency oscillations between collocation points. To suppress non-physical oscillations and enforce local continuous smoothness, we introduce a gradient-enhanced Taylor Loss~\cite{10.1016/j.cma.2022.114823, 10.1016/j.cmpb.2024.108081}. It requires both the physical residual and its spatial gradient to simultaneously approach zero. Mathematically, for an arbitrary collocation point $x$, if we perform a first-order Taylor expansion of the residual in the neighborhood $x+\epsilon$:
	
	\begin{equation}
		R(x+\epsilon) \approx R(x) + \epsilon \cdot \frac{\partial R}{\partial x}
		\label{eq:taylor_exp}
	\end{equation}
	
	To ensure the minimization of the residual $R(x+\epsilon)$ within the neighborhood, we must simultaneously minimize both the zero-order term $R(x)$ and the first-order derivative term $\partial R/\partial x$. Consequently, we define the Taylor Loss $L_{\text{taylor}}$ as the Mean Squared Error (MSE) of the partial derivatives of the physical residuals with respect to the spatial coordinate $x'$:
	
	\begin{equation}
		\mathcal{L}_{\text{taylor}} = \frac{1}{N_c} \sum_{i=1}^{N_c} \left[ | \partial_{x'} R_{\text{cont}}^i |^2 + | \partial_{x'} R_{\text{mom}}^i |^2 \right]
		\label{eq:loss_taylor}
	\end{equation}
	
	where the gradient term of the residual, $\partial R/\partial x'$, is also precisely computed via automatic differentiation techniques. Under the standard physical loss, the model is constrained solely at the collocation points, allowing non-physical high-frequency oscillations to proliferate between them. Conversely, upon introducing $\mathcal{L}_{\text{taylor}}$, an additional spatial gradient constraint is imposed alongside the zero-residual enforcement. This functions as a robust ``neighborhood regularization'' within the solution space, effectively penalizing drastic fluctuations in the residual field and compelling the network to learn a solution that not only satisfies the physical equations at the collocation points but also maintains high physical consistency within its local neighborhoods.
	
	\subsubsection{Adaptive Loss Weighting Strategy}
	The training of PINNs is inherently a multi-objective optimization problem, where the total loss function is composed of data terms, physics terms, and other regularization terms (such as Taylor Loss). Since the gradient magnitudes of different loss components often differ by orders of magnitude and change dynamically during training, using fixed weight coefficients often leads the optimizer to prioritize fitting simpler terms (e.g., boundary conditions) while neglecting complex physical constraints, thereby resulting in entrapment in local optima.
	
	To address this optimization challenge, we explore and adopt a quantum computing-inspired adaptive weighting strategy~\cite{arxiv-2509.20733, 10.3390/e26080649, 10.1088/2632-2153/ad43b2, 10.1088/2632-2153/ada3ab}. The core motivation of this strategy is to simulate the mechanism of tunneling through energy barriers in quantum annealing processes, facilitating the optimizer in escaping local minima. Specifically, instead of relying on manually tuned fixed weights, this algorithm dynamically adjusts the weight $W_k$ for each component based on the gradient statistical properties and descent rates observed throughout the training history~\cite{10.1137/20m1318043, 10.1016/j.jcp.2022.111722}.
	
	This strategy aims to mitigate the gradient imbalance problem in multi-objective optimization by dynamically balancing the gradient contributions of different terms. Computational evaluation indicates that in the low-curvature benchmark geometry (Vessel 0), this method effectively improves the convergence stability of the model. However, when applied to the highly tortuous complex geometry (Vessel 4), solely relying on such algorithmic-level weight adjustments failed to achieve the expected optimization results, and the convergence accuracy of the model remained limited. This result suggests that for such complex problems, merely improving the optimization strategy may be insufficient.
	
	\subsection{Double Correction Physics-Informed Neural Network (DCP-INN) Architecture}
	To address the convergence bottlenecks encountered by conventional algorithmic enhancement strategies under the dual constraints of complex geometry and sparse data, this section presents the core contribution of this study---the proposed DCP-INN framework. This section first theoretically analyzes the failure mechanism of standard PINNs when handling high-frequency geometric features (Section C.1). Subsequently, it details a dual-network structure designed to decouple the flow field baseline from high-frequency residuals (Section C.2), and introduces a specialized regularization mechanism to ensure the physical reliability of the correction process (Section C.3).
	
	\subsubsection{Analysis of Optimization Failure under Complex Geometries}
	When the generic enhancement components (Taylor Loss and adaptive weighting) are applied to the highly tortuous Vessel 4, the model's performance deteriorates significantly, failing to reconstruct the internal velocity field. This optimization failure stems from the coupling of two core factors. First, the inherent spectral bias of deep neural networks prioritizes low-frequency trends~\cite{10.1137/20m1318043, arxiv-1911.08655}, missing the rich high-frequency features (e.g., sharp pressure gradients) induced by complex geometries. Second, sparse observational data exacerbates the non-convexity of the optimization landscape, lacking sufficient spatial anchors to constrain the solution space. Consequently, the optimizer becomes trapped in sub-optimal local minima. Overcoming this bottleneck requires fundamental innovation at the network architecture's inductive bias level rather than merely patching loss functions.
	
	\subsubsection{Architecture Design and Principles of DCP-INN}
\begin{figure*}[!t]
		\centering
		\includegraphics[width=0.85\textwidth]{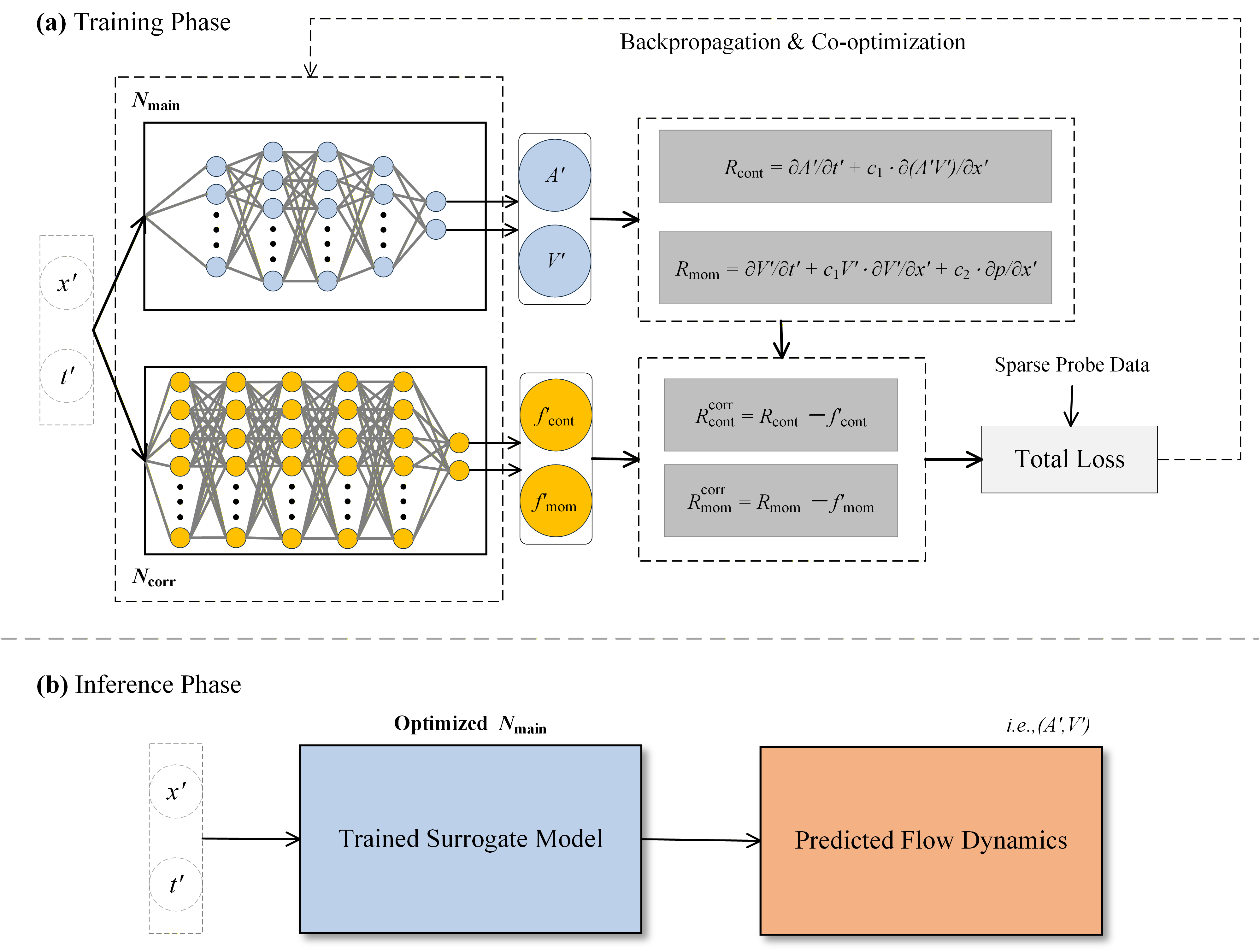}
		\caption{Architecture and information flow of the DCP-INN.
			(a) Training Phase: Non-dimensionalized spatiotemporal coordinates ($x', t'$) are input in parallel. The main network in the upper branch ($N_{\text{main}}$, blue nodes) outputs state variables ($A', V'$); the correction network in the lower branch ($N_{\text{corr}}$, yellow nodes) outputs correction fields ($f_{\text{cont}}', f_{\text{mom}}'$). The grey boxes illustrate the computation of the corrected physical residuals using both outputs, which are ultimately aggregated into the Total Loss for backpropagation and co-optimization.
			(b) Inference Phase: The correction network is detached, and solely the trained $N_{\text{main}}$ independently outputs the Predicted Flow Dynamics.}
		\label{fig:architecture_dcpinn}
	\end{figure*}

	To fundamentally address the aforementioned optimization bottlenecks, we propose the DCP-INN architecture. The core idea of this framework is to adopt a ``divide-and-conquer'' strategy, decoupling the solution task of a single network into two sub-tasks: ``baseline learning'' and ``residual correction''~\cite{10.1016/j.jcp.2019.109020, arxiv-1705.06869, 10.3390/app12188972}. The overall architecture and information flow are illustrated in Fig. \ref{fig:architecture_dcpinn}.
	
	The DCP-INN employs parallel modules: the Main Network ($N_{\text{main}}$) predicts state variables to capture the low-frequency physical baseline, using a Softplus activation on $A'$ to structurally ensure strictly positive cross-sectional areas. Concurrently, the wider and deeper Correction Network ($N_{\text{corr}}$) outputs correction fields $f_{\text{cont}}'$ and $f_{\text{mom}}'$ to capture the high-frequency physical residuals missed by $N_{\text{main}}$ due to spectral bias.
	
	The core mechanism of DCP-INN lies in counteracting the physical errors generated by the Main Network via the correction fields. During the training process, we no longer directly minimize the original physical residuals $R_{\text{cont}}$ and $R_{\text{mom}}$, but instead minimize the modified physical residuals defined as follows:
	
	\begin{equation}
		\mathcal{R}_{i}^{\text{corr}} = \mathcal{R}_{i}(A', V') - f_{i}'(x', t'), \quad i \in \{\text{cont}, \text{mom}\}
		\label{eq:corr_res}
	\end{equation}
	
	This architecture introduces a key training/inference asymmetry. In the training phase, $N_{\text{main}}$ and $N_{\text{corr}}$ are collaboratively updated via the total loss function. $N_{\text{corr}}$ acts as a learnable, spatially-aware slack variable. By absorbing high-frequency errors, it smooths the optimization landscape of the Main Network, enabling it to converge more easily to the correct low-frequency baseline. Conversely, in the inference phase, once training is completed, $N_{\text{corr}}$ is completely discarded. We solely use the fully optimized $N_{\text{main}}$ as the final surrogate model. At this point, the weights of $N_{\text{main}}$ have implicitly encoded the physical constraint information, allowing it to independently perform efficient predictions.
	
	\subsubsection{Correction Field Regularization}
	To prevent $N_{\text{corr}}$ from excessively dominating physical learning and reducing $N_{\text{main}}$ to a mere data fitter, we introduce an Occam's Razor-based regularization loss~\cite{10.1016/j.ins.2016.09.017, 10.1109/tmi.2018.2887072, 10.1109/tmi.2024.3494271, 10.1016/j.patcog.2024.110474, arxiv-2011.13456}. This term enforces sparsity in the correction fields ($f_{\text{cont}}', f_{\text{mom}}'$), compelling the optimizer to leverage $N_{\text{corr}}$ solely in regions where $N_{\text{main}}$ fails (e.g., high-frequency geometric variations). It is defined as the $L_2$ norm of the outputs:
	
	\begin{equation}
		\mathcal{L}_{\text{corr-reg}} = \frac{1}{N_c} \sum_{i=1}^{N_c} \left( |f_{\text{cont}}'(x_i', t_i')|^2 + |f_{\text{mom}}'(x_i', t_i')|^2 \right)
		\label{eq:loss_reg}
	\end{equation}
	
	By incorporating this regularization term into the total loss function, we establish a dynamic balance between ``satisfying physical constraints'' and ``maintaining model parsimony,'' thereby ensuring the physical interpretability and robustness of the final solution.
	
	\subsection{Implementation Details and Evaluation Metrics}
	All model training was implemented using PyTorch~\cite{arxiv-1912.01703} and executed on a single NVIDIA A100 GPU. To ensure reproducibility and delineate the architectural evolution, Table \ref{tab:topology} details the key hyperparameter configurations across the baseline, generic enhanced, and final DCP-INN models.
	
	\begin{table*}[!t]
		\centering
		\caption{Network topology evolution and training hyperparameter configurations of key models. Provides a detailed comparison of core parameters across the baseline PINN, the generic enhanced model, and the final DCP-INN architecture, encompassing the topologies of both the main and correction networks, as well as the loss function compositions.}
		\label{tab:topology}
		\begin{tabular}{llcc}
			\toprule
			\textbf{Model Stage} & \textbf{Main Network ($N_{\text{main}}$)} & \textbf{Correction Net ($N_{\text{corr}}$)} & \textbf{Loss Components} \\
			\midrule
			Baseline PINN & Rectangular [$50\times 4$] & None & $\mathcal{L}_{\text{data}}+\mathcal{L}_{\text{phys}}$ \\
			Enhanced PINN & Rectangular [$50\times 4$] & None & $\mathcal{L}_{\text{baseline}}+\mathcal{L}_{\text{taylor}}$ \\
			DCP-INN (Initial) & Rectangular [$50\times 4$] & Rectangular [$64\times 3$] & $\mathcal{L}_{\text{baseline}}+\mathcal{L}_{\text{taylor}}+\mathcal{L}_{\text{corr-reg}}$ \\
			DCP-INN (Final) & Diamond [50, 100, 100, 50] & Deep \& Wide [$128\times 5$] & $\mathcal{L}_{\text{baseline}}+\mathcal{L}_{\text{taylor}}+\mathcal{L}_{\text{corr-reg}}$ \\
			\bottomrule
		\end{tabular}
	\end{table*}
	
	Through a series of systematic architecture tuning, we established two core network design principles applicable to the double correction framework and applied them to the final model. First, to effectively capture the high-frequency physical residuals induced by complex geometry, the Correction Network ($N_{\text{corr}}$) is designed to possess a larger model capacity than the Main Network, i.e., employing a deeper and wider layer structure. Second, for the Main Network ($N_{\text{main}}$), computational evaluations indicate that an asymmetric ``diamond-shaped'' topology (where layer width first expands and then contracts, e.g., [50, 100, 100, 50]) is superior to the traditional rectangular structure in terms of the balancing efficiency between feature extraction and regression accuracy.
	
	In terms of model performance evaluation, to quantitatively measure the consistency between the reconstructed flow field and the ground truth over the entire spatiotemporal domain, we employ the global relative $L_2$ error as the core metric~\cite{10.1016/j.cmpb.2018.02.001}. Its mathematical definition is as follows:
	
	\begin{equation}
		E_{L2} = \frac{\|u_{\text{pred}} - u_{\text{true}}\|_2}{\|u_{\text{true}}\|_2} = \frac{\sqrt{\sum_{i=1}^N |u_{\text{pred}}^{(i)} - u_{\text{true}}^{(i)}|^2}}{\sqrt{\sum_{i=1}^N |u_{\text{true}}^{(i)}|^2}}
		\label{eq:l2_error}
	\end{equation}
	
	where $u$ represents the vector form of the predicted field, and $N$ is the total number of evaluation points. A lower value of this metric indicates higher reconstruction accuracy of the model for complex hemodynamics.
	
	\section{Results}
	\label{sec:results}

	Based on the data generation workflow described previously, this study instantiates two vessel models with significant geometric differences to comprehensively evaluate the adaptability of the DCP-INN framework under distinct fluid environments. The study subjects, Vessel 0 and Vessel 4, are both derived from the reconstruction of real patient Computed Tomography Angiography (CTA) images~\cite{10.1101/2023.10.19.563201, arxiv-2502.18185, 10.1093/ehjci/ehaa946.0154, 10.1007/s11548-021-02471-5, arxiv-2107.09049}. Specifically, Vessel 0 represents a relatively straight vessel segment. Its comparatively regular geometric features make it suitable as a baseline, utilized to validate the fundamental effectiveness of physical constraint components while excluding the interference of extreme curvature. In contrast, Vessel 4 is a segment of the internal carotid artery characterized by high tortuosity and drastic curvature variations. It represents the core challenge this study aims to address: achieving precise reconstruction in complex flow fields strongly induced by geometry. Although these two models are morphologically distinct, their simulation settings both adhere to strict physical conservation laws. Key parameters, including specific geometric dimensions, fluid physical properties, and Reynolds number ranges, are detailed in Table \ref{tab:parameters}. 
	
	\begin{table*}[!t]
		\centering
		\caption{Physical and geometric parameter configurations of the vessel models. Details the key geometric dimensions, fluid dynamic properties, and non-dimensional reference scales for both the straight (Vessel 0) and highly tortuous (Vessel 4) vessels.}
		\label{tab:parameters}
		\begin{tabular}{llcc}
			\toprule
			\textbf{Parameter Category} & \textbf{Physical Property / Scale} & \textbf{Vessel 0 (Baseline)} & \textbf{Vessel 4 (Complex)} \\
			\midrule
			Geometric Features & Total Length ($L$) & 6.77 cm & 14.17 cm \\
			& Tortuosity Index (TI) & 1.11 (Mild) & 1.60 (Severe) \\
			\addlinespace
			Fluid Properties & Blood Density ($\rho$) & 1.06 g/cm$^3$ & 1.06 g/cm$^3$ \\
			& Dynamic Viscosity ($\mu$) & 0.04 dyn$\cdot$s/cm$^2$ & 0.04 dyn$\cdot$s/cm$^2$ \\
			\addlinespace
			Reference Scales & $L_{\text{ref}}$, $T_{\text{ref}}$ & 6.7730 cm, 0.7143 s & 6.7730 cm, 0.7143 s \\
			& $A_{\text{ref}}$, $V_{\text{ref}}$ & 0.5121 cm$^2$, 51.3389 cm/s & 0.5121 cm$^2$, 51.3389 cm/s \\
			\addlinespace
			Hemodynamics & Peak Reynolds Number (Re) & 1147.934 & 1221.942 \\
			\bottomrule
		\end{tabular}
	\end{table*}
	
	Regarding the construction of the quantitative evaluation system, this study focuses primarily on the global numerical accuracy and predictive fidelity of the flow field reconstruction. To this end, we employ the global relative $L_2$ error ($E_{L_2}$) defined previously to quantify the overall deviation between the predicted flow field and the ground truth, thereby quantitatively measuring the model's capacity to approximate and generalize complex hemodynamics under extremely sparse data constraints.
	
	Utilizing this metric, our initial evaluations reveal the limitations of the Enhanced PINN in handling complex geometries. Although it accurately reconstructs the flow in the low-curvature Vessel 0 (9.70\% relative $L_2$ error), its performance declines on the highly tortuous Vessel 4, with prediction errors reaching 41.49\% (Table \ref{tab:performance}). This geometry-induced degradation delineates the applicability boundary of conventional single-network PINN architectures in real clinical scenarios.
	
	\begin{table}[!t]
		\centering
		\caption{Quantitative performance comparison of model architectures under complex geometry (Vessel 4). Details the error evolution from the standard baseline to the final DCP-INN model. Here, ``Generic Enhanced'' specifically denotes the single-network model integrated with Taylor loss and an adaptive weighting strategy.}
		\label{tab:performance}
		\begin{tabular}{lcc}
			\toprule
			\textbf{Model Stage} & \textbf{Loss Components} & \makecell{\textbf{Relative $L_2$} \\ \textbf{Error (\%)}} \\
			\midrule
			Baseline PINN & $\mathcal{L}_{\text{data}} + \mathcal{L}_{\text{phys}}$ & 33.65 \\
			\addlinespace
			Enhanced PINN & $\mathcal{L}_{\text{baseline}} + \mathcal{L}_{\text{taylor}}$ & 41.49 \\
			\addlinespace
			DCP-INN (Initial) & $\mathcal{L}_{\text{baseline}} + \mathcal{L}_{\text{taylor}} + \mathcal{L}_{\text{corr-reg}}$ & 14.81 \\
			\addlinespace
			DCP-INN (Final) & $\mathcal{L}_{\text{baseline}} + \mathcal{L}_{\text{taylor}} + \mathcal{L}_{\text{corr-reg}}$ & 12.87 \\
			\bottomrule
		\end{tabular}
	\end{table}
	
\begin{figure*}[!t]
		\centering
		\includegraphics[width=0.85\textwidth]{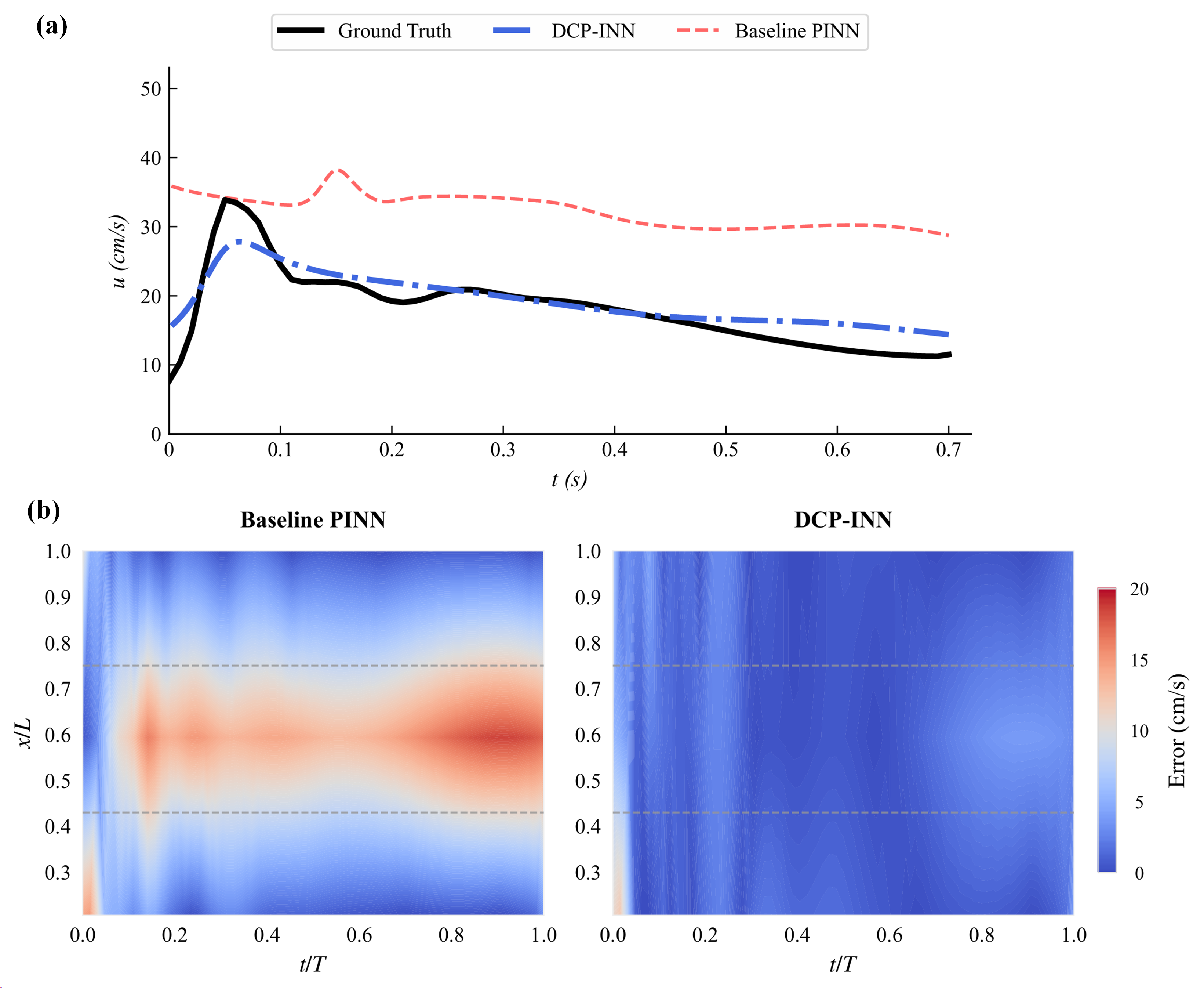}
		\caption{Flow field reconstruction validation and error comparison under complex geometry. (a) Velocity waveform comparison: Single-cardiac-cycle velocity time series extracted at the feature point of maximum curvature in Vessel 4. Black solid lines represent the ground truth, red dashed lines denote the baseline model, and blue dash-dot lines indicate the DCP-INN. (b) Spatial error heatmaps: Distribution of the absolute error $|u_{\text{pred}} - u_{\text{true}}|$ across the full spatiotemporal domain. The two horizontal dashed lines delineate the core spatial interval of high vascular curvature.}
		\label{fig:reconstruction_error}
	\end{figure*}

	To examine the physical mechanisms behind this failure, we performed a spatiotemporal visualization analysis of the baseline model's predictions on Vessel 4. As illustrated by the velocity waveform comparison in Fig. \ref{fig:reconstruction_error}(a), the baseline model's predictions (red dashed line) exhibit typical ``over-smoothing'' characteristics. Although it captures the overall flow trend, it fails to resolve the high-frequency pulsation details induced by sharp vessel curvature. This phenomenon aligns with the inherent spectral bias of deep neural networks, where a single network prioritizes fitting low-frequency components and struggles to capture high-frequency features within limited gradient update steps. In the high-curvature regions of Vessel 4, complex geometric boundaries induce abrupt local velocity gradient changes. During optimization, the network treats these high-frequency physical features as ``noise'' and filters them out, yielding predictions that are smooth yet erroneous. The spatial error heatmap in Fig. \ref{fig:reconstruction_error}(b) further corroborates this observation. The error concentrates in regions with the highest vessel curvature, indicating that a single network cannot effectively deconstruct the high-dimensional nonlinear mapping introduced by complex geometries under the weak constraints of sparse data.
	
	In contrast, the proposed DCP-INN architecture improves reconstruction performance. By decoupling the solution task into ``low-frequency baseline prediction'' and ``high-frequency residual correction,'' DCP-INN overcomes the spectral bias limitations of a single network. Quantitative results show that under identical sparse inputs and training cycles, DCP-INN reduces the reconstruction error on Vessel 4 from 41.49\% (baseline) to 14.81\%~\cite{10.1007/s10554-023-02815-z, 10.1002/mrm.29134}, and further to 12.87\% after final topology tuning. The waveform details in Fig. \ref{fig:reconstruction_error}(a) indicate that the DCP-INN prediction curve (blue dotted line) maintains phase synchronization with the ground truth and accurately reproduces peak and valley variations. Concurrently, the error heatmap in Fig. \ref{fig:reconstruction_error}(b) shows that the high-error regions previously concentrated at curvatures are suppressed, leaving only a sporadic error distribution. This comparative evaluation demonstrates that, rather than merely approximating the ground truth numerically, the DCP-INN architecture enhances the model's capability to capture high-frequency hemodynamic features induced by complex geometries via architectural-level inductive biases.
	
\begin{figure*}[!t]
		\centering
		\includegraphics[width=0.85\textwidth]{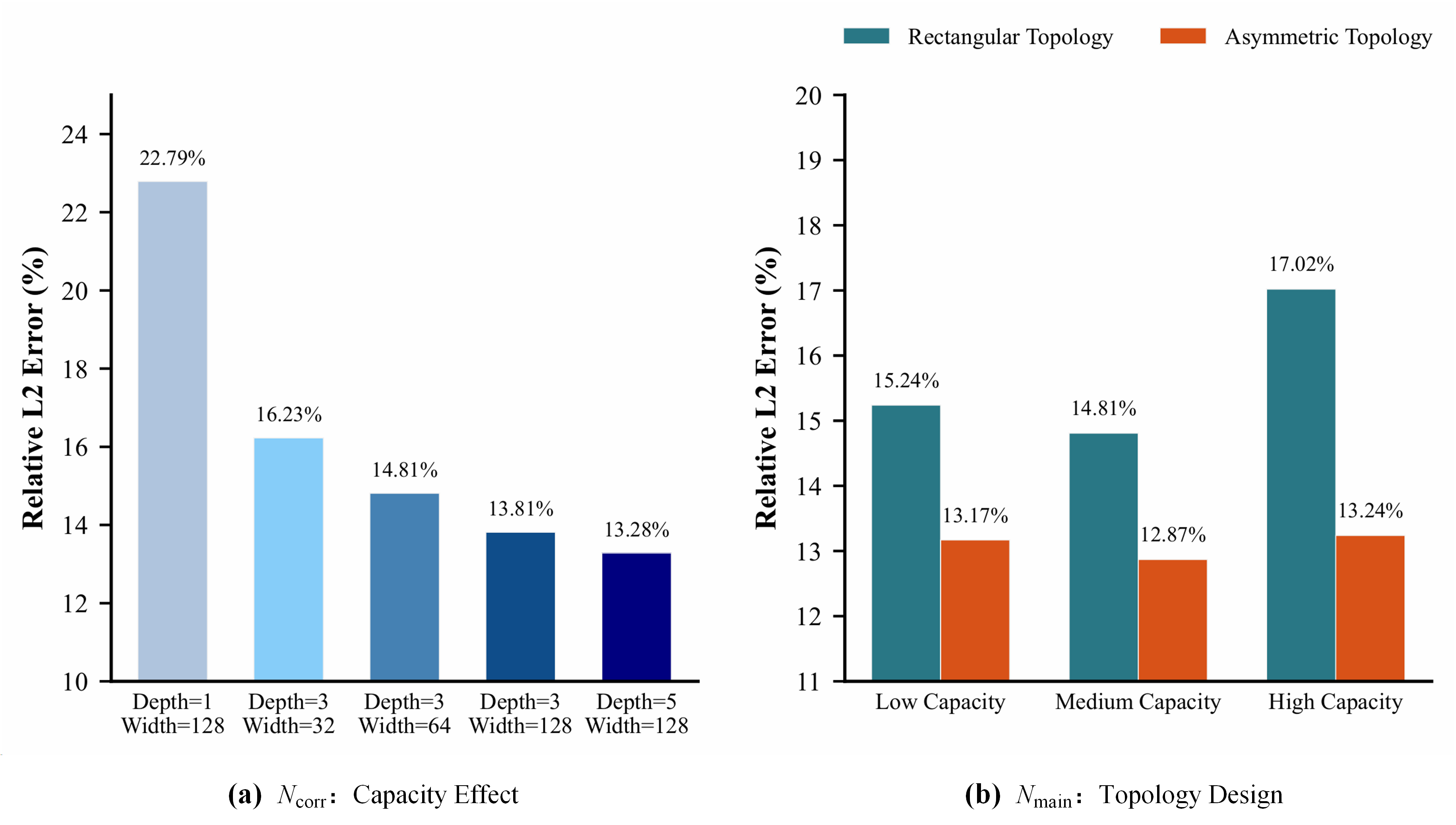}
		\caption{Network topology sensitivity analysis and optimal architecture search. The y-axis denotes the Relative $L_2$ Error (\%). (a) $N_{\text{corr}}$ Capacity Effect: The blue gradient bars display the error variation trends under progressively increasing network depth and width configurations. (b) $N_{\text{main}}$ Topology Design: Compares the error performance of the traditional rectangular topology (teal bars) versus the asymmetric ``diamond-shaped'' topology (orange bars) across low, medium, and high capacity configurations with similar parameter counts.}
		\label{fig:topology}
	\end{figure*}

	The effectiveness of the DCP-INN architecture stems not only from the ``divide-and-conquer'' decoupling strategy but, more fundamentally, relies on aligning the topological capacities of the sub-networks with their specific physical-scale tasks.A systematic sensitivity analysis of the Main Network ($N_{\text{main}}$) and Correction Network ($N_{\text{corr}}$) reveals differences in their feature manifold learning. As shown in Fig. \ref{fig:topology}(a), the reconstruction performance of the Correction Network exhibits a ``capacity threshold effect''. Data indicate that when the network width is below 64 or the depth is less than 3 layers, the model is limited by its fitting capability and struggles to eliminate the high-frequency residuals induced by complex geometries via gradient descent~\cite{10.1016/j.media.2009.07.011}. Conversely, when the width is expanded to 128 and the depth to 5 layers, the model crosses a performance inflection point and enters a stable convergence zone, with the error dropping and stabilizing. This phenomenon quantitatively confirms that capturing high-frequency physical details requires a higher parameter capacity than fitting the low-frequency baseline. In contrast, the Main Network exhibits lower sensitivity to width but demonstrates a structural dependence on topological morphology. As illustrated in Fig. \ref{fig:topology}(b), compared to the traditional rectangular stacked structure, the asymmetric ``diamond-shaped'' topology (i.e., expanding then contracting, e.g., [50, 100, 100, 50]) achieves a balance between feature extraction efficiency and regression stability. This design principle of ``asymmetric capacity allocation''---endowing the Correction Network with high degrees of freedom to capture details while constraining the Main Network to maintain baseline robustness---constitutes the structural foundation for DCP-INN to achieve an error of 12.87\% under complex geometries.
	
	\section{Discussion}
	
	This study addresses a fundamental bottleneck restricting PINNs in complex anatomies: the spectral bias dilemma when processing multi-scale fluid features. While baseline models perform well in straight vessels (Vessel 0), their accuracy degrades in highly tortuous arteries (Vessel 4), constrained by a trade-off between fitting global trends and capturing local high-frequency details. Instead of counteracting this mathematical property, the DCP-INN architecture accommodates it through an architectural decoupling strategy. This framework orthogonally decomposes the complex spatiotemporal flow field: the main network captures low-frequency, large-scale physical trends, while the correction network compensates for high-frequency residuals induced by geometric nonlinearities. Furthermore, the Softplus activation function prevents the numerical collapse commonly observed under extremely sparse data, while the Taylor expansion-based high-order loss propagates local derivative information from the probes to unobserved regions. This integrated strategy of architectural frequency-division and physical regularization reduces the reconstruction error from 41.49\% to 12.87\% and recovers the high-frequency physical textures of the flow field.
	
	At the level of clinical translation, DCP-INN overcomes the inherent trade-off between spatial and temporal resolution in traditional hemodynamic assessments. By integrating the temporal advantages of TCD with the anatomical precision of CTA/MRI, DCP-INN reconstructs a continuous flow field with high-frequency details at any spatial location. Furthermore, this technology demonstrates greater clinical adaptability than traditional computational fluid dynamics (CFD). As a mesh-free method, DCP-INN inherently circumvents the numerical artifacts caused by grid quality. More importantly, it flexibly assimilates sparse downstream probe data as soft constraints, ensuring convergence to physically credible solutions even under unknown or uncertain boundary conditions. Ultimately, this framework enables the rapid construction of personalized digital models using patients' preoperative anatomical structures and baseline blood flow data, providing a low-cost virtual evaluation platform for interventional treatments such as carotid artery stenting (CAS)~\cite{10.1038/s41746-023-00853-4, 10.1016/j.patrec.2017.11.021}.
	
	Despite its clinical potential, this study acknowledges several limitations. First, the ground truth is derived from 1D ROM numerical simulations, which may not fully capture the complex noise distributions and fluid-structure interaction (FSI) effects present in real clinical 4D Flow MRI data. Second, the DCP-INN framework relies on 1D equations, inherently neglecting 3D spatial features such as secondary flow vorticity or non-axisymmetric shear. Third, the model is trained on a single standard cardiac cycle, limiting its capacity to handle heart rate variability or non-periodic flow events. Finally, the physics-constrained optimization process still requires hours of computation. Future work will explore transfer learning strategies to improve computational efficiency and extend this architecture to full 3D flow field reconstruction tasks~\cite{10.1109/tmi.2024.3391651, 10.1109/tmi.2024.3462988, 10.1109/tmi.2025.3594724}.
	
	\section{Conclusion}
	\label{sec:conclusion}
	
	This study proposes the DCP-INN framework to tackle the ill-posed inverse problem of reconstructing hemodynamics from sparse data in complex vascular geometries. To overcome the spectral bias bottleneck of standard PINNs in handling multi-scale fluid features, we employ a synergistic ``main-correction'' dual-network architecture. By processing the flow field across different frequency bands, the DCP-INN framework successfully decouples the low-frequency physical baseline from high-frequency geometric perturbations. Furthermore, by introducing a higher-order loss based on Taylor expansion, the DCP-INN framework significantly enhances the physical consistency and robustness of the model under extremely sparse data constraints. Validated on complex geometries derived from real patient imaging, the DCP-INN framework demonstrates exceptional performance in highly tortuous internal carotid arteries. It drastically reduces the relative reconstruction error from 41.49\% (baseline model) to 12.87\%, successfully recovering complex flow textures often lost by conventional methods. The methodology presented in this study offers a novel clinical tool and approach for research into the development of low-cost, high-resolution personalized cardiovascular digital twins. Beyond this specific application, this study provides compelling evidence for the generalizability of physics-informed deep learning to a wide range of challenging problems in biofluid mechanics and related fields.


\begin{thebibliography}{00}
		\bibitem{10.1016/j.media.2009.07.011}
		D. Lesage \emph{et al.}, ``A review of 3D vessel lumen segmentation techniques: models, features and extraction schemes,'' \emph{Med. Image Anal.}, 2009.
		
		\bibitem{10.1016/j.cmpb.2018.02.001}
		S. Moccia \emph{et al.}, ``Blood vessel segmentation algorithms—Review of methods, datasets and evaluation metrics,'' \emph{Comput. Methods Programs Biomed.}, 2018.
		
		\bibitem{10.1063/5.0259296}
		M. A. Ur Rehman \emph{et al.}, ``Fluid-structure interaction analysis of pulsatile flow in arterial aneurysms with physics-informed neural networks and computational fluid dynamics,'' \emph{Phys. Fluids}, 2025.
		
		\bibitem{10.1002/cnm.3639}
		M. R. Pfaller \emph{et al.}, ``Automated generation of 0D and 1D reduced-order models of patient-specific blood flow,'' \emph{Int. J. Numer. Meth. Biomed. Eng.}, 2022.
		
		\bibitem{10.1016/j.jcp.2018.10.045}
		M. Raissi, P. Perdikaris, and G. E. Karniadakis, ``Physics-informed neural networks: A deep learning framework for solving forward and inverse problems involving nonlinear partial differential equations,'' \emph{J. Comput. Phys.}, 2018.
		
		\bibitem{10.1109/tmi.2022.3161653}
		M. Sarabian, H. Babaee, and K. Laksari, ``Physics-Informed Neural Networks for Brain Hemodynamic Predictions Using Medical Imaging,'' \emph{IEEE Trans. Med. Imag.}, 2022.
		
		\bibitem{10.1109/tmi.2025.3587636}
		J. Kang \emph{et al.}, ``Flow-Rate-Constrained Physics-Informed Neural Networks for Flow Field Error Correction in Four-Dimensional Flow Magnetic Resonance Imaging,'' \emph{IEEE Trans. Med. Imag.}, 2025.
		
		\bibitem{10.1016/j.cmpb.2024.108427}
		A. Sen \emph{et al.}, ``Physics-Informed Graph Neural Networks to solve 1-D equations of blood flow,'' \emph{Comput. Methods Programs Biomed.}, 2024.
		
		\bibitem{arxiv-1911.08655}
		R. Wang \emph{et al.}, ``Towards Physics-informed Deep Learning for Turbulent Flow Prediction,'' \emph{arXiv preprint arXiv:1911.08655}, 2019.
		
		\bibitem{arxiv-1502.05767}
		A. G. Baydin \emph{et al.}, ``Automatic differentiation in machine learning: a survey,'' \emph{J. Mach. Learn. Res.}, 2018.
		
		\bibitem{10.1016/j.cpc.2025.109569}
		Y. Liu \emph{et al.}, ``ICPINN: Integral conservation physics-informed neural networks based on adaptive activation functions for 3D blood flow simulations,'' \emph{Comput. Phys. Commun.}, 2025.
		
		\bibitem{10.1016/j.cma.2022.114823}
		J. Yu \emph{et al.}, ``Gradient-enhanced physics-informed neural networks for forward and inverse PDE problems,'' \emph{Comput. Methods Appl. Mech. Eng.}, 2022.
		
		\bibitem{10.1016/j.cmpb.2024.108081}
		A. Aghaee and M. O. Khan, ``Performance of Fourier-based activation function in physics-informed neural networks for patient-specific cardiovascular flows,'' \emph{Comput. Methods Programs Biomed.}, 2024.
		
		\bibitem{arxiv-2509.20733}
		Y. Huang \emph{et al.}, ``PALQO: Physics-informed Model for Accelerating Large-scale Quantum Optimization,'' \emph{arXiv preprint arXiv:2509.20733}, 2025.
		
		\bibitem{10.3390/e26080649}
		C. Trahan, M. Loveland, and S. Dent, ``Quantum Physics-Informed Neural Networks,'' \emph{Entropy}, 2024.
		
		\bibitem{10.1088/2632-2153/ad43b2}
		A. Sedykh \emph{et al.}, ``Hybrid quantum physics-informed neural networks for simulating computational fluid dynamics in complex shapes,'' \emph{Mach. Learn.: Sci. Technol.}, 2024.
		
		\bibitem{10.1088/2632-2153/ada3ab}
		A. Setty, R. Abdusalamov, and F. Motzoi, ``Self-adaptive physics-informed quantum machine learning for solving differential equations,'' \emph{Mach. Learn.: Sci. Technol.}, 2025.
		
		\bibitem{10.1137/20m1318043}
		S. Wang, Y. Teng, and P. Perdikaris, ``Understanding and Mitigating Gradient Flow Pathologies in Physics-Informed Neural Networks,'' \emph{SIAM J. Sci. Comput.}, 2021.
		
		\bibitem{10.1016/j.jcp.2022.111722}
		L. D. McClenny and U. M. Braga-Neto, ``Self-adaptive physics-informed neural networks,'' \emph{J. Comput. Phys.}, 2022.
		
		\bibitem{10.1016/j.jcp.2019.109020}
		X. Meng and G. E. Karniadakis, ``A composite neural network that learns from multi-fidelity data: Application to function approximation and inverse PDE problems,'' \emph{J. Comput. Phys.}, 2019.
		
		\bibitem{arxiv-1705.06869}
		Y. Yang \emph{et al.}, ``ADMM-Net: A Deep Learning Approach for Compressive Sensing MRI,'' \emph{IEEE Trans. Pattern Anal. Mach. Intell.}, 2018.
		
		\bibitem{10.3390/app12188972}
		M. Shafiq and Z. Gu, ``Deep Residual Learning for Image Recognition: A Survey,'' \emph{Appl. Sci.}, 2022.
		
		\bibitem{10.1016/j.ins.2016.09.017}
		Y. Son and J. Lee, ``Active learning using transductive sparse Bayesian regression,'' \emph{Inf. Sci.}, 2016.
		
		\bibitem{10.1109/tmi.2018.2887072}
		K. C. Tezcan \emph{et al.}, ``MR Image Reconstruction Using Deep Density Priors,'' \emph{IEEE Trans. Med. Imag.}, 2018.
		
		\bibitem{10.1109/tmi.2024.3494271}
		S. Han \emph{et al.}, ``Physics-informed Score-based Diffusion Model for Limited-angle Reconstruction of Cardiac Computed Tomography,'' \emph{IEEE Trans. Med. Imag.}, 2024.
		
		\bibitem{10.1016/j.patcog.2024.110474}
		G. Kim \emph{et al.}, ``Depth-aware guidance with self-estimated depth representations of diffusion models,'' \emph{Pattern Recognit.}, 2024.
		
		\bibitem{arxiv-2011.13456}
		Y. Song \emph{et al.}, ``Score-Based Generative Modeling through Stochastic Differential Equations,'' \emph{Int. Conf. Learn. Represent.}, 2020.
		
		\bibitem{arxiv-1912.01703}
		A. Paszke \emph{et al.}, ``PyTorch: An Imperative Style, High-Performance Deep Learning Library,'' \emph{Adv. Neural Inf. Process. Syst.}, 2019.
		
		\bibitem{10.1101/2023.10.19.563201}
		H. Narotamo, M. Silveira, and C. A. Franco, ``3DVascNet: an automated software for segmentation and quantification of vascular networks in 3D,'' \emph{bioRxiv}, 2023.
		
		\bibitem{arxiv-2502.18185}
		A. Iltaf \emph{et al.}, ``VesselSAM: Leveraging SAM for Aortic Vessel Segmentation with LoRA and Atrous Attention,'' \emph{arXiv preprint arXiv:2502.18185}, 2025.
		
		\bibitem{10.1093/ehjci/ehaa946.0154}
		A. Chandrashekar \emph{et al.}, ``A deep learning approach to automate high-resolution blood vessel reconstruction on computerised tomography images with or without the use of contrast agents,'' \emph{Eur. Heart J.}, 2020.
		
		\bibitem{10.1007/s11548-021-02471-5}
		J. Wang \emph{et al.}, ``Coarse-to-fine multiplanar D-SEA UNet for automatic 3D carotid segmentation in CTA images,'' \emph{Int. J. Comput. Assist. Radiol. Surg.}, 2021.
		
		\bibitem{arxiv-2107.09049}
		L. Chen \emph{et al.}, ``Deep Open Snake Tracker for Vessel Tracing,'' \emph{arXiv preprint arXiv:2107.09049}, 2021.
		
		\bibitem{10.1007/s10554-023-02815-z}
		D. Long \emph{et al.}, ``Super-resolution 4D flow MRI to quantify aortic regurgitation using computational fluid dynamics and deep learning,'' \emph{Int. J. Cardiovasc. Imaging}, 2023.
		
		\bibitem{10.1002/mrm.29134}
		G. S. Roberts \emph{et al.}, ``Virtual injections using 4D flow MRI with displacement corrections and constrained probabilistic streamlines,'' \emph{Magn. Reson. Med.}, 2021.
		
		\bibitem{10.1038/s41746-023-00853-4}
		K. Sel \emph{et al.}, ``Physics-informed neural networks for modeling physiological time series for cuffless blood pressure estimation,'' \emph{npj Digit. Med.}, 2023.
		
		\bibitem{10.1016/j.patrec.2017.11.021}
		C. Li \emph{et al.}, ``Generative adversarial dehaze mapping nets,'' \emph{Pattern Recognit. Lett.}, 2019.
		
		\bibitem{10.1109/tmi.2024.3391651}
		G. Ruan \emph{et al.}, ``Magnetic Resonance Electrical Properties Tomography Based on Modified Physics-Informed Neural Network and Multiconstraints,'' \emph{IEEE Trans. Med. Imag.}, 2024.
		
		\bibitem{10.1109/tmi.2024.3462988}
		Z.-X. Cui \emph{et al.}, ``Physics-Informed DeepMRI: k-Space Interpolation Meets Heat Diffusion,'' \emph{IEEE Trans. Med. Imag.}, 2024.
		
		\bibitem{10.1109/tmi.2025.3594724}
		M. Morik \emph{et al.}, ``Enhancing Brain Source Reconstruction by Initializing 3D Neural Networks with Physical Inverse Solutions,'' \emph{IEEE Trans. Med. Imag.}, 2025.
		
	\end{thebibliography}
\end{document}